\documentclass[10pt,aps]{revtex4}
\usepackage{amssymb}
\usepackage{latexsym}
\usepackage{epsfig}
\usepackage{float}
\usepackage{graphicx,epsfig, color}
\usepackage{graphicx}
\usepackage{graphics}
\usepackage{subfigure}
\usepackage{amsmath}
\usepackage{epstopdf} 
\usepackage{float}
\usepackage{mathrsfs}

\begin{document}
	\title{\textbf{Tensor condensate accompanied by chiral transition in a strong magnetic field}}
	\author{Yuan-Hao Niu$^{1}$, Xin-Jian Wen$^{1,2}$}
	\address{
		$^{1}$Institute of Theoretical Physics , Shanxi University, Taiyuan 030006, Shanxi, China\\
		$^{2}$Collaborative Innovation Center of Extreme Optics, Shanxi University, Taiyuan 030006, Shanxi, China}

	\begin{abstract}
		We investigate tensor condensates and chiral condensates in the (2+1)-flavor Nambu-Jona-Lasinio model at finite temperature and density in the presence of a strong magnetic field. The emergence of the tensor condensate is attributed to the four-fermion interaction. It is shown that a sufficiently large chemical potential is necessary for the occurrence of a phase transition towards tensor condensate. Furthermore, we investigate the correlation between tensor condensate and spin polarization, which accounts for the oscillatory behavior of the spin polarization during the transition from the chiral condensate to tensor condensate phase.
	\end{abstract}
	
	\maketitle
	\section{introduction}
	As our understanding of matter has deepened, the investigation of the phase diagram of strong interactions has emerged as a fundamental and crucial topic. The behavior of Quantum Chromodynamics (QCD) matter in magnetic fields contributes to our understanding of nuclear matter under extreme conditions, such as early universe evolution \cite{Vachaspati:1991nm}, Relativistic Heavy Ion Collider experiments \cite{Skokov:2009qp, Deng:2012pc}, neutron stars \cite{Kurkela:2022elj}, and compact stars \cite{CamaraPereira:2016chj, Baym:2017whm}. Additionally, the spin polarization of quark matter could be pertinent to various astrophysical phenomena, such as neutron stars \cite{Tsue:2012nx}, spontaneous polarization of quark matter \cite{Maruyama:2017mqv}, and vacuum polarization with ferromagnetism \cite{Maruyama:2000cw, Tsue:2016age, Tsue:2012jz}. In theoretical research, first-principles calculations, particularly lattice Quantum Chromodynamics (LQCD), offer reliable predictions of QCD phase properties and transitions at zero baryon chemical potential and finite temperature \cite{Fukushima:2010bq}. These predictions serve as a valuable reference in confirming the chiral phase transitions and spin-polarization phase transitions under magnetic fields. In Relativistic Heavy Ion Collider (RHIC) experiment, a magnetic field as intense as $10^{15}G$ was observed \cite{Deng:2012pc}. Neutron stars are known to have strong magnetic fields on their surface and in the core \cite{Kurkela:2022elj, Watts:2016uzu}. It is noted that in neutron stars with intense magnetic fields, one fascinating phenomenon is the occurrence of stable pulses and the emergence of multiple magnetic poles \cite{Riley:2019yda}. Therefore, our focus is on the spin polarization of QCD matter and the provide explanations for these phenomena.\par 
	In the study of spin polarization, the Nambu-Jona-Lasinio(NJL) model can be employed by including the four-fermion interaction. In the relativistic limit, the expression of spin polarization is consistent with that given by the four-fermion interaction 	\cite{Ferrer:2010wz, Paulucci:2010uj}. The introduction of a strong magnetic field in the magnetic NJL(MNJL) model presents us with a multitude of new challenges \cite{Menezes:2008qt, Menezes:2009uc, Avancini:2011zz, Ferrer:2013noa}. First, a strong magnetic field will cause energy level splitting for charged particles. Second, the spin of particles will precess in an external magnetic field.\par 
	In 1976, 't Hooft introduced quantum effects involving four-dimensional pseudoscalar interactions, which were fundamental in subsequent studies on spin polarization \cite{tHooft:1976snw}. In the 1900s, the relativistic framework was considered more suitable for investigating spin polarization, which was implicitly defined through the spin polarization operator \cite{Koch:1987py, Niembro:1990tc}. Subsequently, it was revealed that the final outcome of Lagrangian for spinors, as calculated by 't Hooft in Ref \cite{Bohr:2012mg, tHooft:1976snw}. In 2000, Toshitaka investigated the ferromagnetism of quark fluid and demonstrated that quark matter could form a ferromagnetic phase due to single gluon exchange interactions, which exhibits a so-called spin flip effect \cite{Tatsumi:1999ab}. Since 2012, the spin-polarized phase has gained much attention due to the separation of the axial vector term and tensor term in the framework of quark meson model. Bohr found an anomalous magnetic moment phenomenon within the axial vector \cite{Bohr:2012mg}. In 2017, Maruyama suggested that spin polarization results from the contribution of tensor interactions, thereby diminishing the influence of the axial vector \cite{Maruyama:2017mqv}. In 2020, Kagawa incorporated the hexaquark interaction, challenging previous conclusions regarding  tensor interaction terms \cite{Kagawa:2020vsf}. Recently, Qiu and Feng delved into the comparison of anomalous magnetic moments with spin polarization under strong magnetic fields, which offered a method to account for the inverse magnetic catalysis \cite{Qiu:2023kwv}. Additionally the deconfined phase with restored chiral symmetry also had been studied~\cite{Bao:2024glw}. It is interesting that quarkyonic phase could be affected by TSP. Furthermore it had been observed that the Gell-Mann-Oakes-Renner relation breaks down when considering AMM or TSP effect~\cite{Lin:2022ied}.\par
	This work is organized as follows: In Sec.\ref{sec:model}, we present the thermodynamics involving chiral and tensor condensate in a strong magnetic field. In Sec.\ref{sec:calculation}, the numerical result is shown for the phase of tensor condensate, quark chiral condensate, and spin polarization at finite temperature and chemical potential. A summary and short conclusion are presented in Sec.\ref{sec:summeraize}.\par

	\section{THE (2+1)-FLAVORS NJL MODEL UNDER A MAGNETIC FIELD}\label{sec:model}
	
	We consider the following generalized Lagrangian density of the (2+1)-flavors NJL model in a strong magnetic field \cite{tHooft:1976snw, Buballa:1996tm},\par
	\begin{align}
		{\mathcal{L}}={}& {\mathcal{L}_0}+{\mathcal{L}_S}+{\mathcal{L}_T}+{\mathcal{L}_I} -\frac{1}{4} F_{\mu \nu} F^{\mu \nu}, \\
		{\mathcal{L}_0}={}& \bar{\psi}(i{\gamma}^{\mu}{\mathcal{D}}_{\mu} - \hat{m}_0)\psi, \\
		{\mathcal{L}_S}={}& G_s\sum_{a=0}^{8} \big[ (\bar{\psi}\lambda_a \psi)^2+(\bar{\psi}i \gamma^5 \lambda_a \psi)^2\big], \\
		{\mathcal{L}_T}={}& G_t\sum_{a=0}^{8} \big[ (\bar{\psi} \varSigma_3 \lambda_a \psi)^2+(\bar{\psi} \varSigma_3 i \gamma^5 \lambda_a \psi)^2\big], \\
		{\mathcal{L}_I}={}& G_k[det\bar{\psi}(1+\gamma_5)\psi+det\bar{\psi}(1- \gamma_5)\psi],
	\end{align}
	where $\mathcal{L}_I$ is Kobayashi-Maskawa't Hooft term or the determinant interaction term, which leads to the six-point interaction between quarks in the three-flavor case. $\mathcal{L}_T$ is closely related to the spin-spin interaction, which causes spin-polarization condensate. $\mathcal{L}_S$ is four-point tensor interaction between quarks in the three-flavor case, which preserves chiral symmetry \cite{Buballa:2003qv, Tsue:2012nx, Abhishek:2018usv}. $\psi=(u,d,s)^{\top}$ is three-flavor quark field, and the diagonal current quark mass matrix is $\hat{m}_0=\mathrm{diag}_f(m_u, m_d, m_s)$. We adapt color indices $c=(r,b,g)$ and flavor indices $f=(u,d,s)$. $\lambda_a,\:(a=1,...,N_f^2-1)$ 
	are the flavor $\mathrm{SU}(3)$ groups of generators, also known as the Gell-Mann matrix. $\varSigma_3 = \begin{bmatrix}
		\sigma_z & 0 \\ 0 & -\sigma_z
	\end{bmatrix}$ is spin project operator, and related to the Dirac matrix by $\gamma^1 \gamma^2 = i \varSigma_3$. \par
	The significant disparity in masses between the s quark and the u, d quarks gives rise to a violation of the $\mathrm{SU}(3)$ symmetry. The covariant derivative is defined as $\mathcal{D}_{\mu}=\partial_{\mu}+iQA_{\mu}^{ext}$, with the quark charge matrix, $Q=diag(q_u,q_d,q_s)=diag(\frac{2}{3},-\frac{1}{3},-\frac{1}{3})$, and $A_{\mu}^{ext} = (0, 0, Bx,0)$ is the gauge field with the constant magnetic field should point at the z-direction \cite{Qiu:2023kwv, Noronha:2007wg}. In this paper, we adopt the mean-field calculation scheme. The chiral term arises from the four-fermion interaction. In the three-flavor context, a novel interaction term appears, as a six-point interaction. The tensor term characterizes the tensor component of the spin polarization, assuming a spin polarization direction along the z-axis.\par
	In this context, the quark-antiquark condensates are denoted as $\phi_f=\langle\bar{\psi}\psi\rangle$, and the spin polarization condensates, denoted as $F_a=-G_t\langle\bar{\psi}\varSigma_z\lambda_a \psi\rangle$. If the spin polarization parameter is defined as $ \kappa_S = \frac{\rho_{\uparrow} - \rho_{\downarrow}}{\rho_{\uparrow} + \rho_{\downarrow}}$, it can be related to $F_a$ through a Lorentz transformation \cite{Maruyama:2000cw, Tatsumi:1999ab, Bohr:2012mg}. This suggests that the four-fermion interaction derived by 't Hooft can effectively account for spin polarization. In the case of (2+1) flavors of quarks, the Gell-Mann matrix enables the expression of $F_3 = -G_t\langle \bar{\psi}\varSigma_z\lambda_3\psi\rangle$ and $F_8 = -G_t\langle\bar{\psi}\varSigma_z\lambda_8\psi\rangle$ from $F_a$. It is important to note that $F_3$ and $F_8$ do not directly represent spin-up or spin-down states, but rather the clustering of quark matter in flavor space. Additionally, the signs of the $F_3$ and $F_8$ may depend on the chosen representation. In this paper, we have chosen the chiral Weyl representation. Meanwhile, we can transform the spin polarization term into a tensor condensate as follow \cite{Bohr:2012mg, Kagawa:2020vsf}, \par
	\begin{align}
		F_u=&{}F_3+\frac{1}{\sqrt{3}}F_8, \\
		F_d=&{}-F_3+\frac{1}{\sqrt{3}}F_8, \\
		F_s=&{}-\frac{2}{\sqrt{3}}F_8. 
	\end{align}
	To obtain the case in the magnetic field, we use the following substitution, from the Landau level and the dispersion relation for the quark \cite{Fraga:2008qn}, $p_0 \to i (\omega_\nu-i\mu_f), \mathbf{p}^2 \to p_z^2+(2n+1-\eta)$, where the spin quantum number is $\eta=\pm1$, and the landau quantum number $n=0,1,2...$. The integral should be changed as $\frac{1}{(2\pi)^4}\int \mathrm{d}^4\mathbf{p} \to \frac{Tq_fB_m}{(2\pi)^2}\sum_{\nu=\pm} \sum_{n=0}^{\infty}\int\mathrm{d}p_z $. In the above relationship, we use $\omega_\nu=(2\nu+1)\pi T$ to represent the Matsubara frequencies. Following the standard scheme \cite{Qiu:2023kwv}, the single-particle energy in the presence of a magnetic field can be expressed as, \par
	\begin{equation}
		E_{f,k,\eta} = \sqrt{p_z^2 + (\sqrt{M_f^2 + 2k|q_fB|} + \eta F_f)^2},
	\end{equation}
	where $M_f$ is the effective mass of quark, $F_f$ is the quark of tensor condensates, $q_f$ is quark electric charge, and $k = 2n+1-\eta$ is the quantum number of Landau level, $p_z$ is the momentum in the z-direction.
	The grand canonical ensemble thermodynamic potential with the chiral and spin polarization condensate is given by \cite{Tatsumi:1999ab},
	\begin{align}
		\Omega(T,\mu,M_u,M_d,M_s,F_3,F_8)={}&-\frac{N_c}{4\pi^2}\sum_{k=0}^{\infty}\sum_{f=u,}^{d,s}\sum_{\eta= \pm} |q_f B|\int\mathrm{d}p_z \biggl\{ 
		E_{f,k=0}+E_{f,k,\eta} \notag \\
		&+T\ln{\big[1+\exp{\big(-\frac{E_{f,k,\eta}-\mu}{T}\big)}\big]}+T\ln{\big[1+\exp{\big(-\frac{E_{f,k,\eta}+\mu}{T}\big)}\big]} \biggr\} \notag \\
		&+\sum_{f=u,}^{d,s}2 G_s \phi_f^2+\frac{F_3^2 + F_8^2}{2G_t}+4 G_k\phi_u \phi_d \phi_s ,
	\end{align}
	where $N_c=3$ is the number of colors, $\mu$ is chemical potential. And we assume all chemical potentials to be equal for simplifying our calculations, $T$ is temperature. Finally, we can formulate the mean-field Lagrangian in a strong magnetic field \cite{Menezes:2008qt},
	\begin{equation}
		\mathcal{L}=\bar{\psi}(iD_{\mu}\gamma^{\mu}-\hat{M}- F_3 \varSigma_3 \lambda_3 - F_8 \varSigma_8 \lambda_8 )\psi -2G_s (\phi_u^2+\phi_d^2+\phi_s^2)+4G_k \phi_u \phi_d \phi_s-\frac{F_3^2}{2G_t}-\frac{F_8^2}{2G_t},
	\end{equation}
	where $\hat{M} = \mathrm{diag}(M_u, M_d, M_s)$, and $M_f$ is the effective quark mass. In preceding calculations, we ignored the electromagnetic interaction terms $\mathcal{L}_M = -\frac{1}{4} F_{\mu \nu} F^{\mu \nu}$, as would not contribute to a normalized thermodynamic potential $\Omega_{eff} = \Omega - \Omega_0 $, where $\Omega_0$ represents the vacuum contribution. \par
	By minimizing thermodynamic potential of the system, we can have the gap equations.
	\begin{align}
		\frac{\partial\Omega}{\partial\phi_f} = \frac{\partial\Omega}{\partial F_3} = \frac{\partial\Omega}{\partial F_8} = 0. \label{eq12}
	\end{align}
	One can also formulate the Fermi distribution function of quarks, 
	\begin{align}
		N_{f,k,\eta}^{\pm} = &{} \frac{1}{1 + \exp[\frac{E_{f,k,\eta} \mp \mu}{T}]} ,
	\end{align}
	and give rise to the net density of quarks, 
	\begin{align}
		\rho_{f} = &{} \sum_{k=0}^{\infty} \alpha_k \frac{N_c|q_f B|}{4 \pi^2} \int dp (N_{f,k,\eta}^+ - N_{f,k,\eta}^-) , 
	\end{align}
	where the degeneracy is $\alpha_k=\delta_{0,k}+\Delta(k)\sum_{\eta=\pm }$, if $k=0$, then $\Delta(0)=0$, else $\Delta(k)=1$. By solving the gap equations, we can determine the effective quark mass,
	\begin{align}
		M_u =&{} m_{u} - 4 G_s \phi_u + 2 G_k \phi_d \phi_s, \\
		M_d =&{} m_{d} - 4 G_s \phi_d + 2 G_k \phi_u \phi_s, \\
		M_s =&{} m_{s} - 4 G_s \phi_s + 2 G_k \phi_d \phi_u, 
	\end{align}
	where $m_f$ is quark current mass. For the purpose of convenience, the thermodynamic potential is divided into a free part and an interacting part. Due to the divergence caused by vacuum contribution, a sharp cutoff $\Lambda$ is employed to guarantee the convergence of the integral. Hence, the total thermodynamic potential is a sum of following three terms,
	\begin{align}
		\Omega_f=&{}\Omega_f^{vac}+\Omega_f^{med}+\Omega^{Int}, 
	\end{align}
	\begin{align}
		\Omega_f^{vac}=&{} -\frac{N_c|q_f B|}{4\pi^2} \int_{\Lambda_z} \,\ \mathrm{d}p_z \bigg[ \sum_{k=0}^{k^{mag+}_{max}}E_{f,k}^+ +\sum_{k=1}^{k^{mag-}_{max}}E_{f,k}^-\bigg] \Theta(p_{\perp}^2)\Theta(\Lambda^2-p_{\perp}^2), \label{eq:(18)} 
	\end{align}
	\begin{align}
		\Omega_f^{med}=&{}-\frac{N_c|q_f B|}{4\pi^2}\int\mathrm{d}p_3 \biggl\{\sum_{k=0}^{k^{\mu}_{max}}\bigg[ T\ln\big[1+\exp(-\frac{E_{f,k}^+ -\mu}{T})\big]+T\ln\big[1+\exp(-\frac{E_{f,k}^+ +\mu}{T})\big] + \notag \\ 
		&\sum_{k=1}^{k^{\mu}_{max}}\bigg[ T\ln\big[1+\exp(-\frac{E_{f,k}^- -\mu}{T})\big]+T\ln\big[1+\exp(-\frac{E_{f,k}^- +\mu}{T})\big] \bigg]\biggr\} , \\
		\Omega_f^{Int}=&{}\sum_{f=u,}^{d,s}2 G_s \phi_f^2+\frac{F_3^2 + F_8^2}{2G_t}
		- 4 G_k \phi_u \phi_d \phi_s,
	\end{align}
	where $\Omega_f^{vac}$ is the quark condensate vacuum dependent on the magnetic field, and $\Theta$ is Heaviside step function. In the view of non-renormalization of the NJL model \cite{Klevansky:1992qe}, the regularization has also been suggested \cite{Kohyama:2016fif, Kohyama:2015hix}. The other parameters are $p_{\perp}^2=(\sqrt{M_f^2+2k|q_f B_m|}+F_f)^2-M_f^2$ , and the Landau levels can be written as,
	\begin{align}
		 \Lambda_z^2=&{}\Lambda^2 - (2k|q_fB_m|-2M_fF_f+2\eta F_f\sqrt{M_f^2+2k|q_fB_m|}) \notag \\
		 k^{\pm}_{max} = &{} \mathrm{Floor} \big[ \frac{(\Lambda \mp F_f)^2 - M_f^2}{2|q_f B|} \big]
	\end{align}
    If $F_3$ approaches the limit 0, the eq.(\ref{eq:(18)}) can be reduced to the form given in Ref \cite{Menezes:2008qt}. \par
    The entropy of quark also can be written as $S_f = - \frac{\partial \Omega_f}{\partial T}$,
    \begin{align}
    	S_f = &{} -\frac{N_c|q_f B|}{4\pi^2} \int \mathrm{d}p_z \biggr\{ \sum_{k=0}^{k^{\mu}_{max}} \big[N^{+}_{f, k,\eta = +1}\ln[N^{+}_{f, k,\eta = +1}] + (1-N^{+}_{f, k,\eta = +1})\ln(1-N^{+}_{f, k,\eta = +1}) \notag \\
    	&{}+ [N^{+}_{f, k,\eta = +1} \leftrightarrow N^{-}_{f, k,\eta = +1}]\big] \notag \\
    	&{} + \sum_{k=1}^{k^{\mu}_{max}} \big[N^{+}_{f, k,\eta = -1}\ln[N^{+}_{f, k,\eta = -1}] + (1-N^{+}_{f, k,\eta = -1})\ln(1-N^{+}_{f, k,\eta = -1}) \notag \\
    	&{}+ [N^{+}_{f, k,\eta = -1} \leftrightarrow N^{-}_{f, k,\eta = -1}]\big] \biggr\},
    \end{align}
	Correspondingly, the corresponding chiral condensate is a sum of two terms as $\phi_f=\phi_f^{vac}+\phi_f^{med}$,
	\begin{align}
		\phi_f^{vac}=&{}-\frac{N_c|q_f B|}{4\pi^2} \int_{\Lambda_z} \,\ \mathrm{d}p_z \big[\sum_{k=0}^{k_{max}} \frac{M_f(\sqrt{M_f^2+2 k|q_f B_m|}+ F_f)}{E_{f,k}^+ \sqrt{M_f^2+2 k|q_f B_m|}} +\sum_{k=1}^{k_{max}} \frac{M_f(\sqrt{M_f^2+2 k|q_f B_m|}- F_f)}{E_{f,k}^- \sqrt{M_f^2+2 k|q_f B_m|}}\big], \\
		\phi_f^{med}=&{} \frac{N_c|q_f B|}{4\pi^2}\int\mathrm{d}p_z \biggr\{ \sum_{k=0}^{k^{\mu}_{max}}\frac{M_f(\sqrt{M_f^2+2k|q_f B|}+F_f)}{E_{f,k}^+ \sqrt{M_f^2 + 2k|q_f B|}}\big[\frac{1}{1+\exp(\frac{E_{f,k}^+ -\mu}{T})}+\frac{1}{1+\exp(\frac{E_{f,k}^+ +\mu}{T})}\big] + \notag \\ 
		&\sum_{k=1}^{k^{\mu}_{max}}\frac{M_f(\sqrt{M_f^2+2k|q_f B|}-F_f)}{E_{f,k}^- \sqrt{M_f^2 + 2k|q_f B|}}\big[\frac{1}{1+\exp(\frac{E_{f,k}^- -\mu}{T})}+\frac{1}{1+\exp(\frac{E_{f,k}^- +\mu}{T})}\big]\biggr\}, 
	\end{align}
	and the spin polarization condensate is also a sum of two terms $FFt_f=FFt_f^{vac}+FFt_f^{med}$,
	\begin{align}
		FFt_f^{vac}=&{}-\frac{N_c|q_f B|}{4\pi^2}\int_{\Lambda_z} \,\ \mathrm{d}p_3 \big[ \sum_{k=0}^{k_{max}} \frac{\sqrt{M_f^2+2 k|q_f B_m|}+ F_f}{E_{f,k}^+ }-\sum_{k=1}^{k_{max}}\frac{\sqrt{M_f^2+2 k|q_f B_m|}- F_f}{E_{f,k}^- }\big], \\
		FFt_f^{med}=&{} \frac{N_c|q_f B|}{4\pi^2}\int\mathrm{d}p_3 \biggr\{ \sum_{k=0}^{k^{\mu}_{max}}\frac{\sqrt{M_f^2+2k|q_f B|}+F_f}{E_{f,k}^+}\big[\frac{1}{1+\exp(\frac{E_{f,k}^+ -\mu}{T})}+\frac{1}{1+\exp(\frac{E_{f,k}^+ +\mu}{T})}\big] - \notag \\ 
		&\sum_{k=1}^{k^{\mu}_{max}}\frac{\sqrt{M_f^2+2k|q_f B|}-F_f}{E_{f,k}^-}\big[\frac{1}{1+\exp(\frac{E_{f,k}^- -\mu}{T})}+\frac{1}{1+\exp(\frac{E_{f,k}^- +\mu}{T})}\big]\biggr\},
	\end{align}
	from the relations $\frac{\partial \Omega_f}{\partial F_3} = \frac{\partial \Omega_f}{\partial F_f} \frac{\partial F_f}{\partial F_3}$ and $\frac{\partial \Omega_f}{\partial F_8} = \frac{\partial \Omega_f}{\partial F_f} \frac{\partial F_f}{\partial F_8}$ in eq.(\ref{eq12}). Then, the following terms could be obtained,
	\begin{align}
		F_3 =&{} G_t(FFt_u - FFt_d), \\
		F_8 =&{} \frac{G_t}{\sqrt{3}}(FFt_u + FFt_d - 2FFt_s) .
	\end{align} 
	Gap equations may have multiple solutions, but the solution with the lowest thermodynamic potential is considered as the physically solution.\par

	\section{NUMERICAL RESULTS AND DISCUSSIONS}\label{sec:calculation}
	For simplifying calculations, we adopt the coupling constants as $\Lambda^2 G_s=1.835, G_t=2G_s, \Lambda^5G_k=9.29$, and a sharp cutoff $\Lambda=631.4 \;\mathrm{MeV}$ \cite{Hatsuda:1994pi, Kohyama:2016fif}. Quark current masses $m_u=m_d=5.6 \;\mathrm{MeV}, m_s=135.7 \;\mathrm{MeV} $ is used to produce the empirical values $f_{\pi}=91 \;\mathrm{MeV}, m_{\pi} =138 \;\mathrm{MeV}, m_K=495.7 \;\mathrm{MeV}, m_{\eta'}=957.5 \;\mathrm{MeV}$. \par
	\subsection{Condensate Versus Temperature}
\begin{figure}[htbp]
		\centering
		\includegraphics[width=8cm]{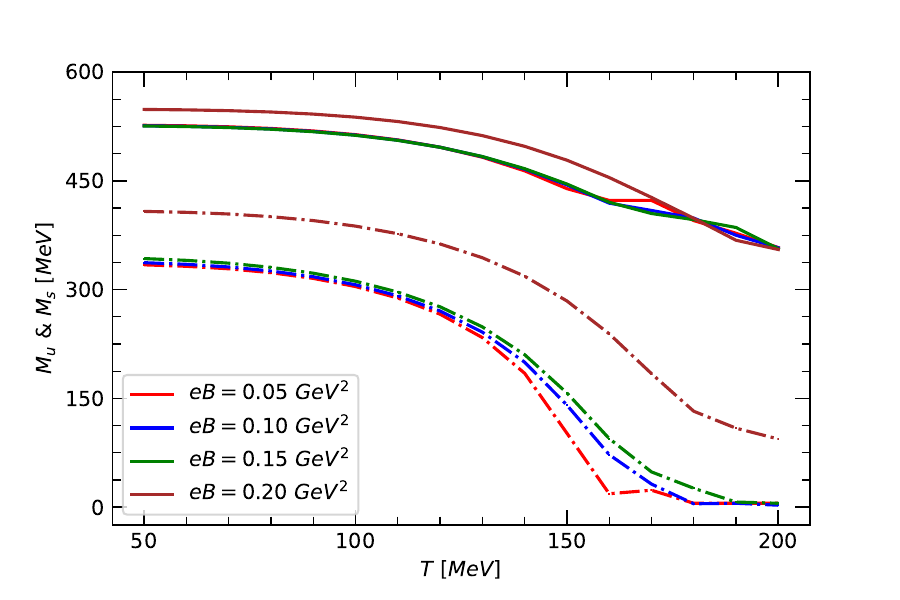}
		\caption{The temperature dependence of the effective masses ($M_u$ and $M_s$) for u and s quarks is shown under four different magnetic fields. The effective masses of u and s quarks are marked by the dotted and solid lines respectively. The various colored lines indicate different magnetic fields: red for $eB=0.05 \;\mathrm{GeV}^2$, blue for $eB=0.10 \;\mathrm{GeV}^2$, green for $eB=0.15 \;\mathrm{GeV}^2$, and brown for $eB=0.20 \;\mathrm{GeV}^2$}
		\label{FIG.1}
	\end{figure}

	FIG. \ref{FIG.1} illustrates the effective mass of u quark and s quark as functions of increasing temperature at a chemical potential $\mu=150 \;\mathrm{MeV}$. The light quark mass $M_u$ exhibits a apparent drop in the low-temperature region, which indicates partial restoration of chiral symmetry. At the magnetic field $eB=0.20 \;\mathrm{GeV}^2$, the spin polarization is enhanced, and results in a partial restoration of chiral symmetry as the temperature increases. The changes in magnetic field have weak effect on the transition temperature of chiral transitions at the temperature larger than $T = 140~\mathrm{MeV}$. Quark matter characterized by tensor condensation displays a magnetic catalytic effect, which is available without the tensor condensation~\cite{Pagura:2016pwr, Bali:2012zg, DElia:2018xwo}. \par
	
	\par

\begin{figure}[H]
		\centering
		\includegraphics[width=12cm]{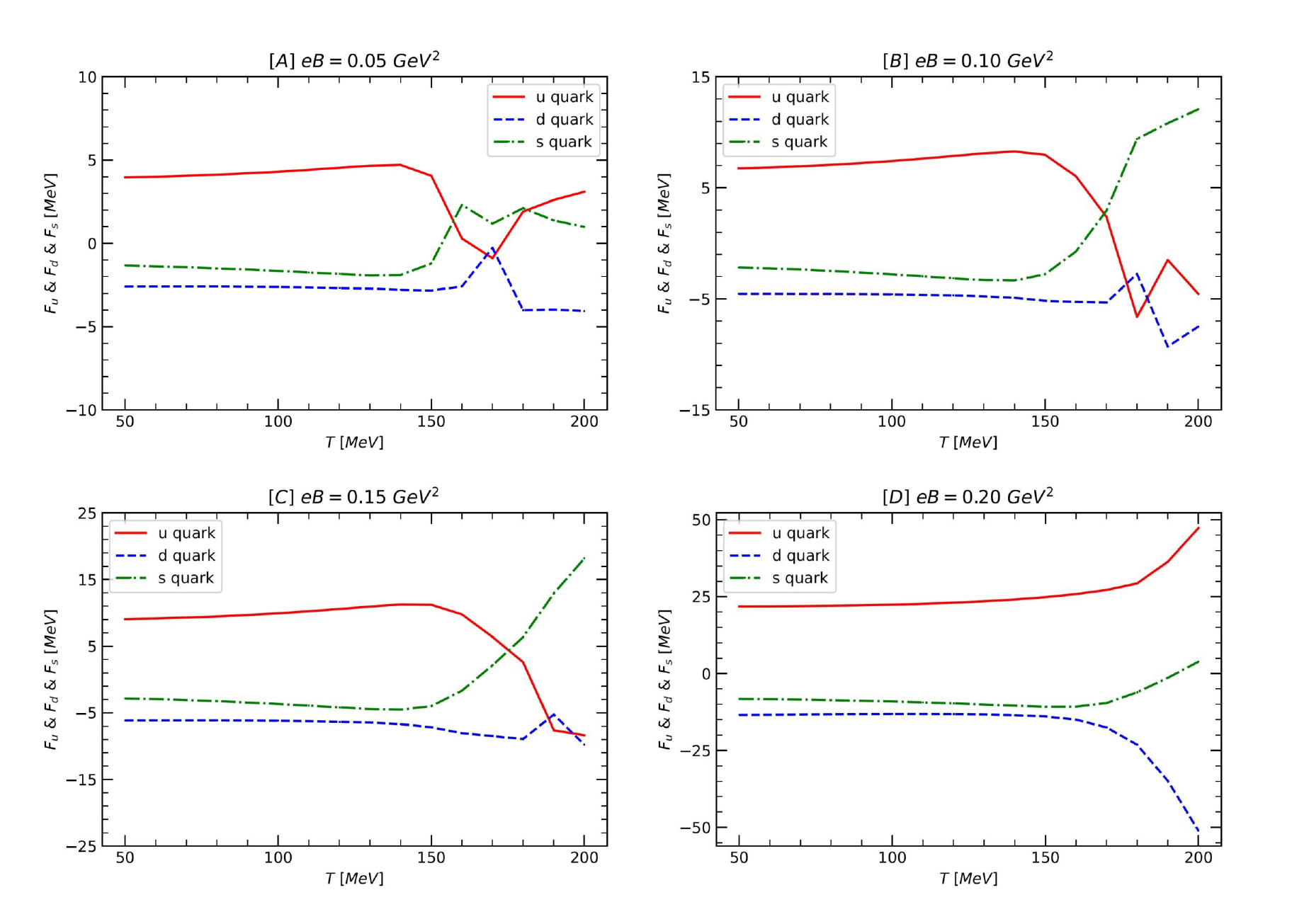}
		\caption{The temperature dependences of the tensor condensation is temperature-dependent for four different magnetic fields, as shown in Panels A ($eB=0.05 \;\mathrm{GeV}^2$), B ($eB=0.10\;\mathrm{GeV}^2$), C ($eB=0.15\;\mathrm{GeV}^2$), and D ($eB=0.20\;\mathrm{GeV}^2$). The red, green, and blue lines denote u, d and s quarks respectively.}
		\label{FIG.2}
	\end{figure}

	Almost horizontal line are observed for the u and s quarks in FIG. \ref{FIG.2}, at magnetic field below $eB=0.20 \;\mathrm{GeV}^2$. $F_3$ can be considered as a measure of the tensor condensate, defined as $F_3= \rho(\uparrow)-\rho(\downarrow)$, which represents the relative mismatch between spin-up and spin-down states densities. Moreover, the absence of $F_3$ for the s quark can be attributed to mathematical calculations involving Gell-Mann matrices. It is concluded that s quarks may exhibit arbitrary spinning directions. In fact, at the low chemical potential and temperature, a tensor condensate transition of quarks is difficult to occur. \par
	
	\par
	\begin{figure}[H]
		\centering
		\includegraphics[width=12cm]{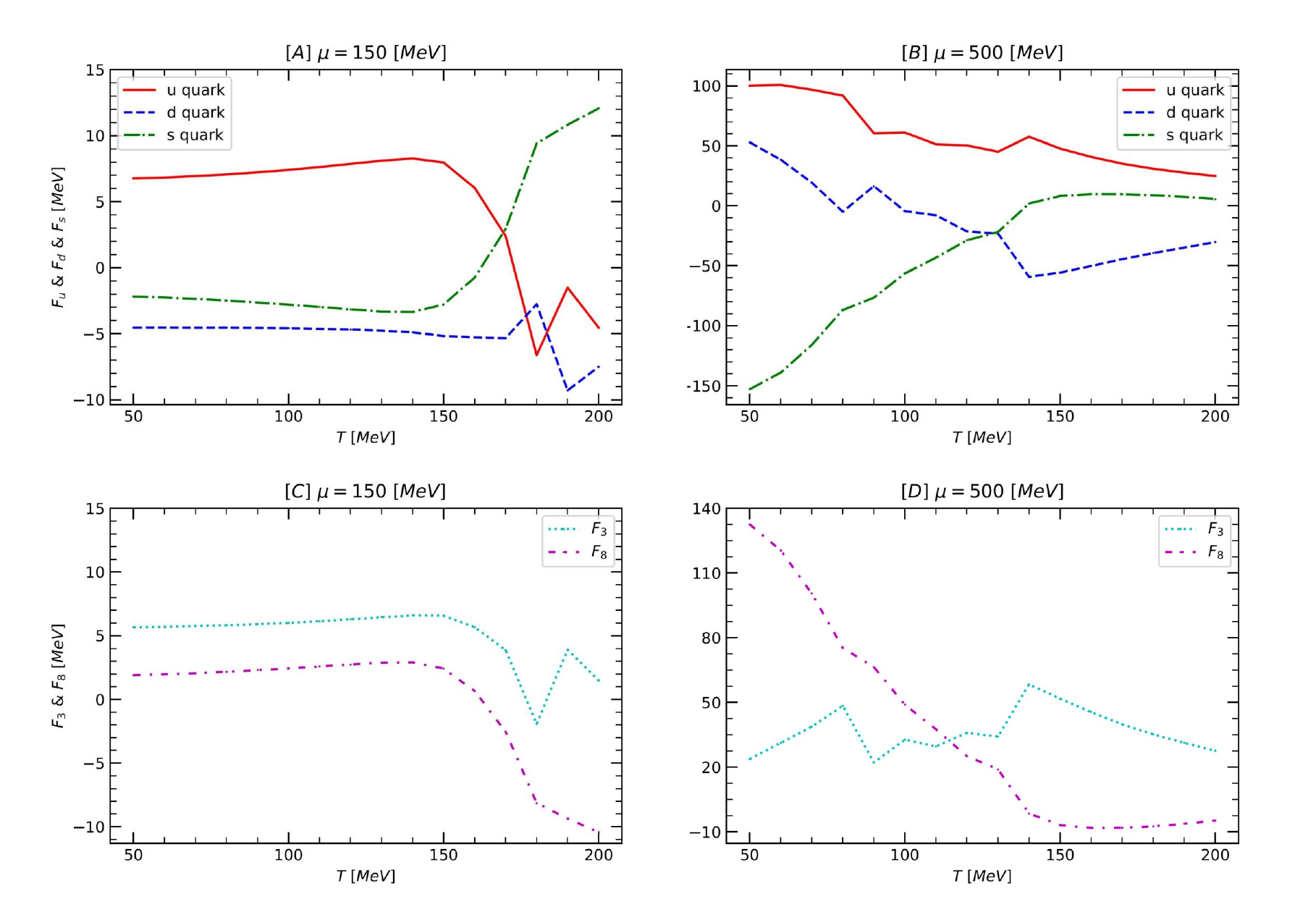}
		\caption{Comparison of tensor condensation and spin polarization at chemical potentials $\mu=150 \;\mathrm{MeV}$ (A, C) and $\mu=500 \;\mathrm{MeV}$ (B, D) in a magnetic field $eB=0.10 \;\mathrm{GeV}^2$. In Panels A and B, the color red represents the u quark, blue represents the d quark, and green represents the s quark. In Panels C and D, purple corresponds to $F_3$, while cyan corresponds to $F_8$.}
		\label{FIG.3}
	\end{figure}\par
	
	The different chemical potentials of tensor condensation are illustrated in FIG. \ref{FIG.3}. Comparing panels (A) and (B), it can be observed that tensor condensation leads to a crossover transition as the temperature increases. In FIG.\ref{FIG.3} panel D, the $F_8$ exhibits a monotone decrease, while $F_3$ shows fluctuations. At a high chemical potential $\mu=500 \;\mathrm{MeV}$ and low temperature $T=50 \;\mathrm{MeV}$, there are significant constraints that render the quark spin aligns with the direction of the magnetic field. As the temperature increases, the thermal motion of quark disrupts the states, which is indicated by the fluctuation of $F_3$. As for $F_8$, its behavior resembles that of ferromagnetic. However, tensor condensation does not exist in the region of low temperature and low chemical potential due to the minimal constraints on spin polarization. \par

	\subsection{Condensate Versus Chemical Potential}
\begin{figure}[htbp]
		\centering
		\includegraphics[width=8cm]{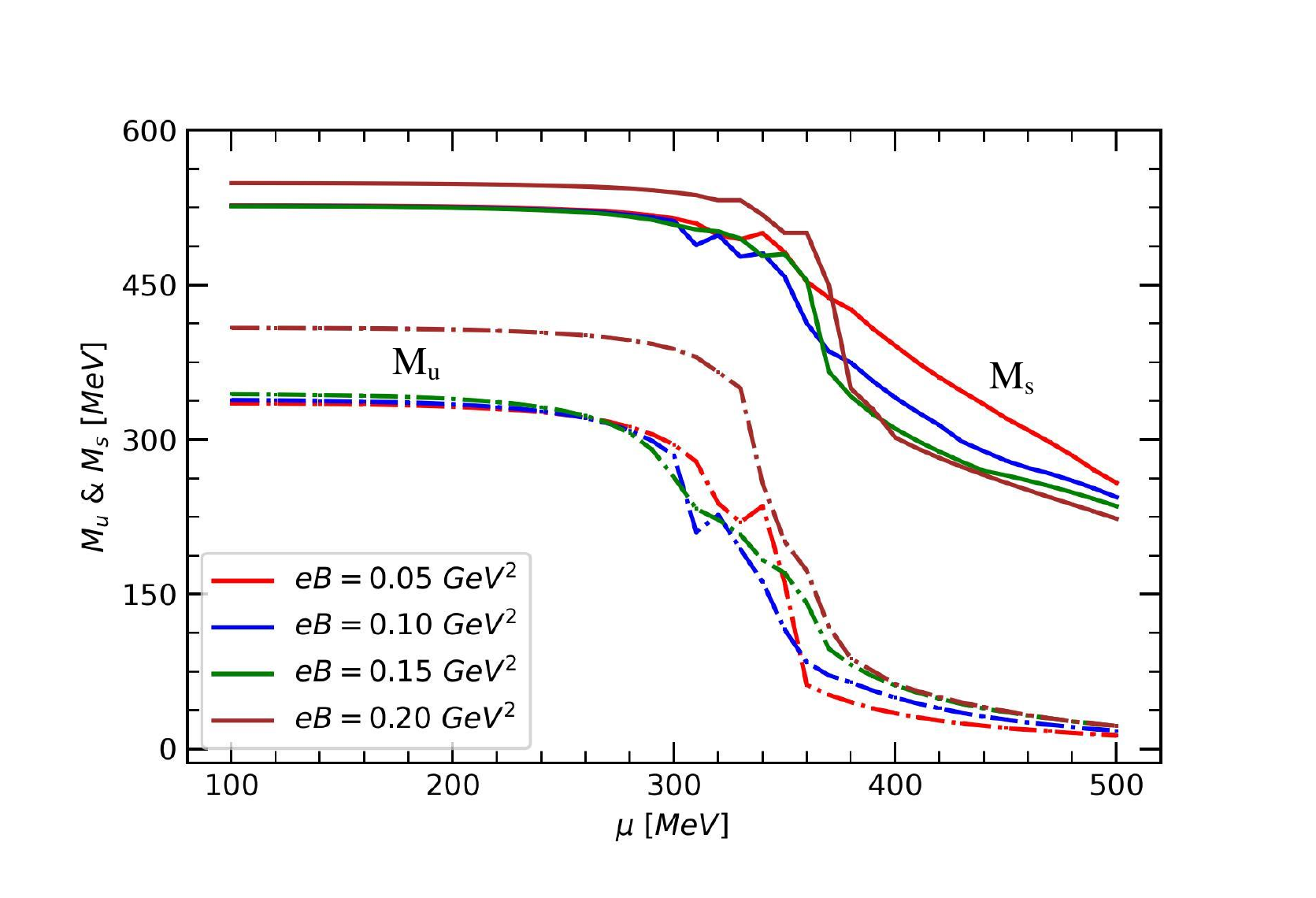}
		\caption{The dependence of the effective masses of u and s quarks on the chemical potential is shown under four different magnetic fields. The dotted line represents the effective mass of the u quark, while the solid line represents the effective mass of the s quark. The colored lines correspond to different magnetic fields: red for $eB=0.05 \;\mathrm{GeV}^2$, blue for $eB=0.10 \;\mathrm{GeV}^2$, green for $eB=0.15 \;\mathrm{GeV}^2$, and brown for $eB=0.20 \;\mathrm{GeV}^2$.}
		\label{FIG.4}
	\end{figure}

	The effective mass of the u and s quarks at four different magnetic fields are depicted as functions of chemical potentials at the temperature $T=50 \;\mathrm{MeV}$ in FIG. \ref{FIG.4}. It is observed that the first-order chiral restoration occurs with an increase in the magnetic field. This can be explained by the appearance of quark tensor condensates. Particularly, at a weak magnetic field $eB=0.05\;\mathrm{GeV}^2$, there is a partial restoration of quark chiral symmetry, which signals a crossover process. At high temperature the effective mass is unequal to current mass in the region of tensor condensation. This discrepancy is attributed to the energy gaining from spin polarization, which compensates for the loss of energy caused by the reduction in the chiral condensate.\par
\begin{figure}[htbp]
		\centering
		\includegraphics[width=15cm]{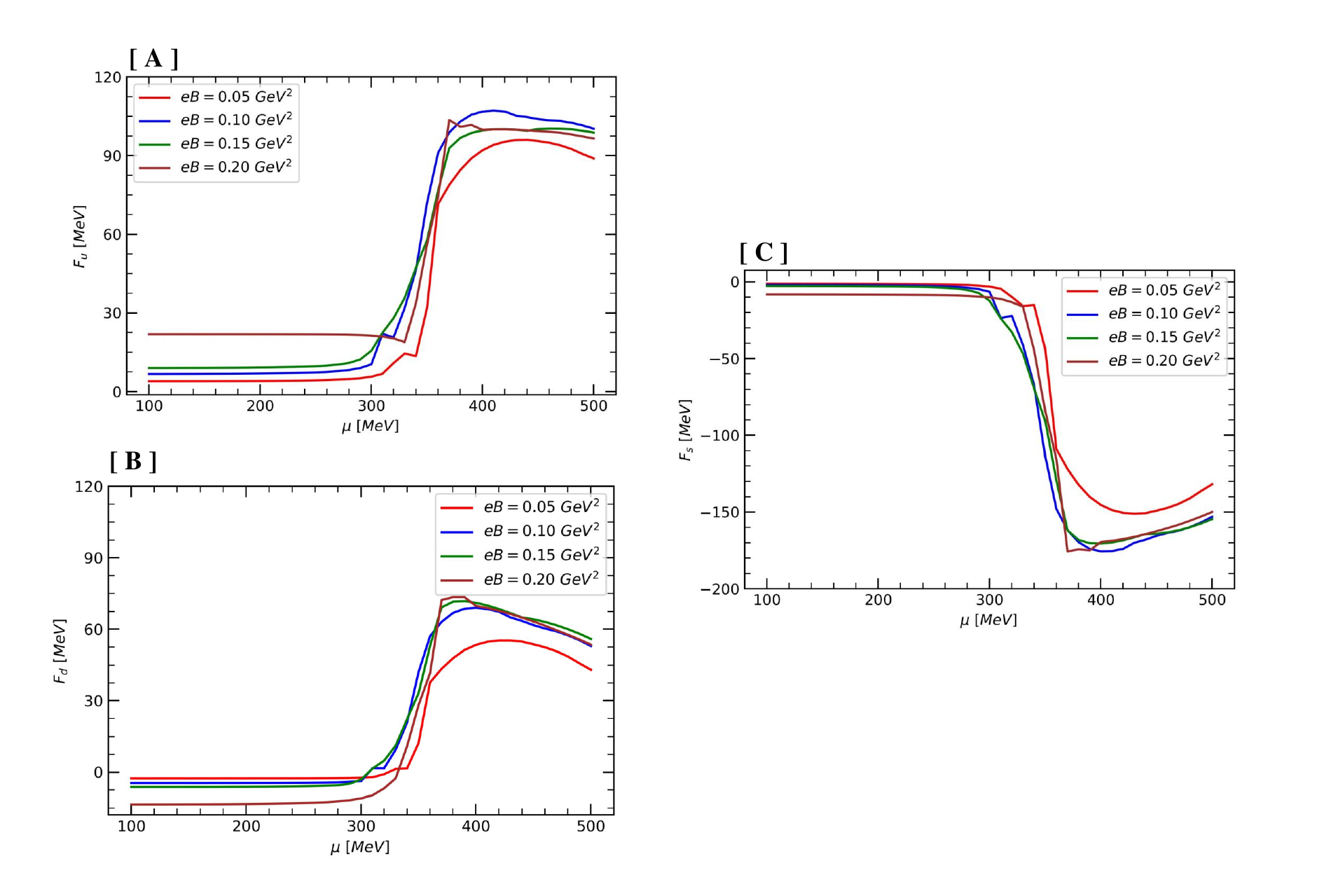}
		\caption{The dependence of quark tensor condensation on the chemical potential is shown for four different magnetic fields. In Panels A, B, and C, are for the u, d, and s quarks respectively.}
		\label{FIG.5}
	\end{figure}

	The tensor condensates at four different magnetic fields is depicted in FIG. \ref{FIG.5} at the temperature $T=50 \;\mathrm{MeV}$. It is evident that, except for a magnetic field of $eB=0.05 \;\mathrm{GeV}^2$ with a crossover process, the tensor condensation demonstrates a first-order transition process at all other magnetic fields. Furthermore, it can be observed that as the magnetic field increases, the crossover region gradually shortens and eventually turns into a first-order transition. In contrast to the chiral condensate shown in FIG. \ref{FIG.4}, the critical chemical potential $\mu=310 \;\mathrm{MeV}$ for the tensor condensation remains relatively unaffected by the magnitude of the magnetic field. This result is consistent with previous studies \cite{Matsuoka:2018huq}.\par
	\par
	\par
	
	\begin{figure}[htbp]
		\centering
		\includegraphics[width=15cm]{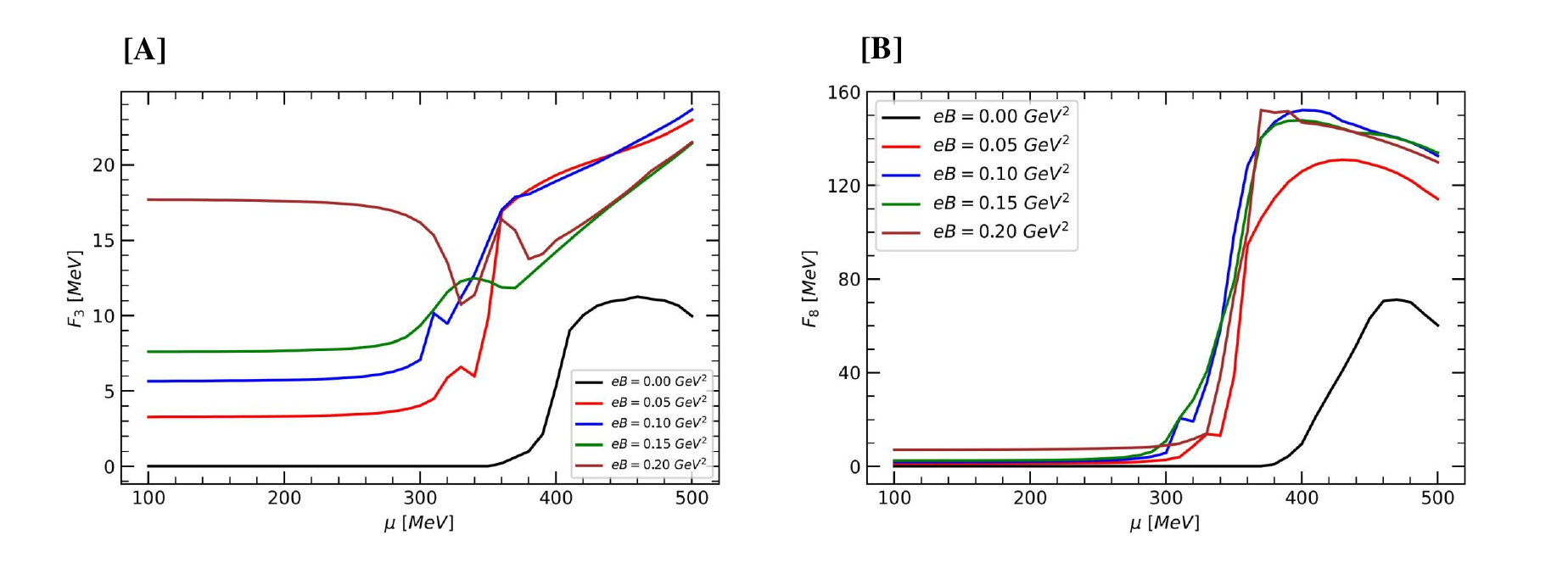}
		\caption{The dependence of quark SP ($F_3$ and $F_8$) on chemical potential ($\mu$) is investigated for four different magnetic fields. In Panel A, $F_3$ is shown, and in Panel B, $F_8$ is shown at different magnetic fields red for $eB=0.05 \;\mathrm{GeV}^2$, blue for $eB=0.10 \;\mathrm{GeV}^2$, green for $eB=0.15 \;\mathrm{GeV}^2$, and brown for $eB=0.20 \;\mathrm{GeV}^2$, and black for $eB=0.00 \;\mathrm{GeV}^2$, represented by colored lines from bottom to top.}
		\label{FIG.6}
	\end{figure}\par
	
	In FIG.\ref{FIG.6}, the behaviors of $F_3$ and $F_8$ under different magnetic field strengths are shown at $T=50 \;\mathrm{MeV}$. It is evident that $F_3$ is significantly enhanced by the magnetic field at the chemical potential below the critical value $\mu=340 \;\mathrm{MeV}$. As the chemical potential continues to increase, the growth rate of $F_3$ at $eB=0.10 \;\mathrm{GeV}^2$ is faster than the case of $eB=0.05 \;\mathrm{GeV}^2$. In the right panel, $F_8$ exhibits a significant increase until it reaches a peak. Further increase of the chemical potential will result in a decrease in $F_8$. This behavior can be interpreted as the onset of quark ferromagnetism occurring at the critical chemical potential. Subsequently, the quark spin aligns with the magnetic field directions. The ferromagnetic properties of quark matter can be accounted for by the phenomenon of tensor condensation in quark matter. The result at $eB=0$ from Ref~\cite{Abhishek:2018usv} is marked by the black solid line for comparison. It is found that the behavior at weak magnetic field is close to the result of ~\cite{Abhishek:2018usv}  The presence of an external magnetic field enhances spin polarization and makes it appear in a larger value at lower chemical potential by comparison with the zero magnetic field.\par
	
	At the larger chemical potential, quarks would generally be in Cooper pairs leading to color superconductivity. It is possible that the tensor condensate diminishes with the formation of bosons, and the spin anti-parallel pairs could lead to the fluctuation of $F_3$.\par
	
\begin{figure}[H]
		\centering
		\includegraphics[width=15cm]{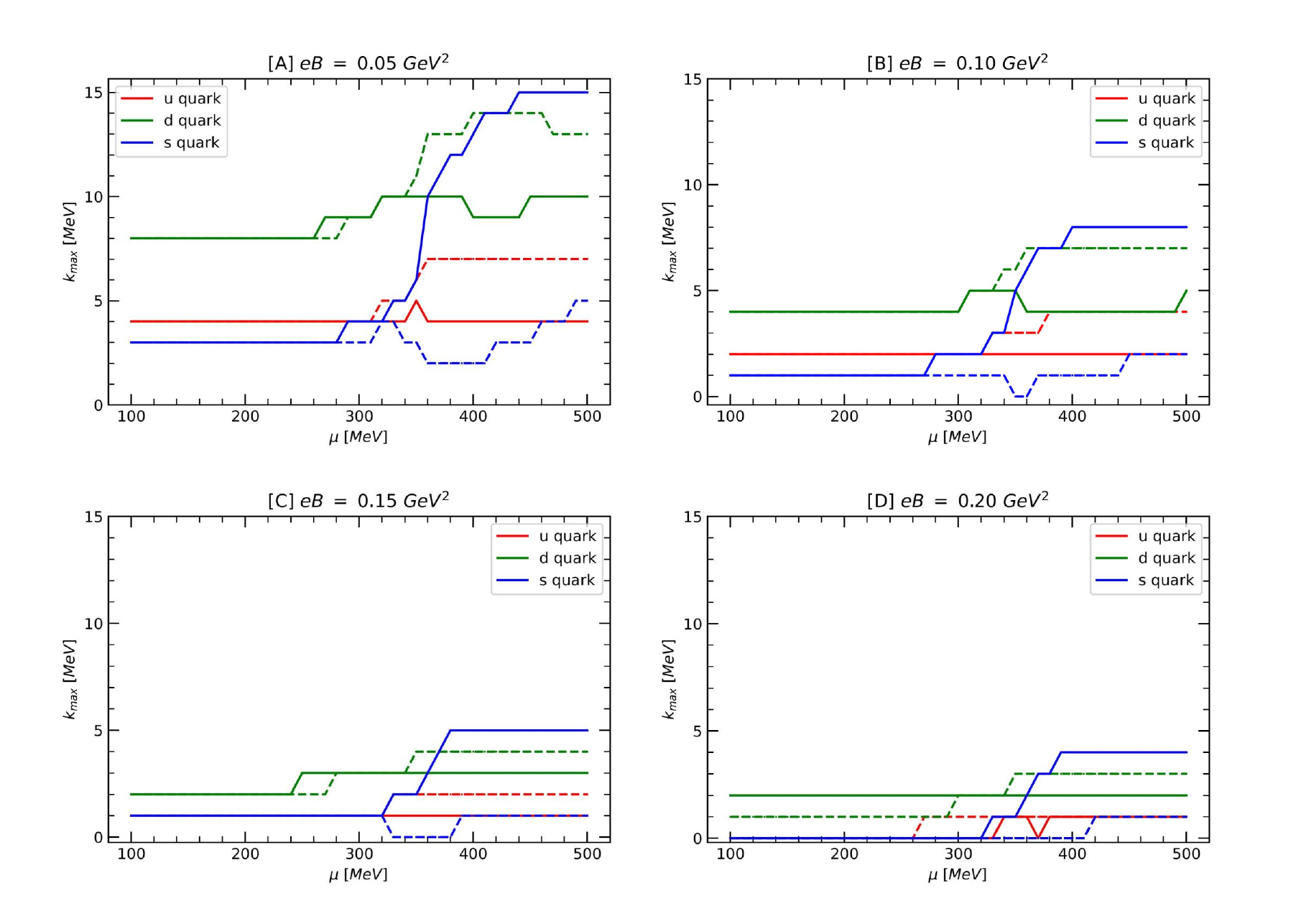}
		\caption{The Landau levels is chemical potential-dependent for four different magnetic fields. The solid line and dashed lines the energy level of $\eta = 1$, and $\eta = -1$, respectively. In Panels A ($eB=0.05 \;\mathrm{GeV}^2$), B ($eB=0.10\;\mathrm{GeV}^2$), C ($eB=0.15\;\mathrm{GeV}^2$), and D ($eB=0.20\;\mathrm{GeV}^2$). The red, green, and blue lines denote u, d and s quark respectively.}
		\label{FIG.7}
	\end{figure}

	Fig.\ref{FIG.7} shows the number of the Landau energy levels of u, d and s quarks at different magnetic fields. With the increase of magnetic fields, Landau energy level decreases. As quarks from the tensor condensate, the Landau energy level increases. The number of $\eta = 1$ energy levels of light quarks is higher than the number of $\eta = -1$ energy levels in tensor condensate. It means that the effect of spin polarization is mainly reflected in the Landau energy level. Of cause, at the oscillation of $F_3$ , the Landau energy level changes accordingly.\par
	
	\par

	\subsection{The map of entropy}
\begin{figure}[H]
	\centering
	\includegraphics[width=15cm]{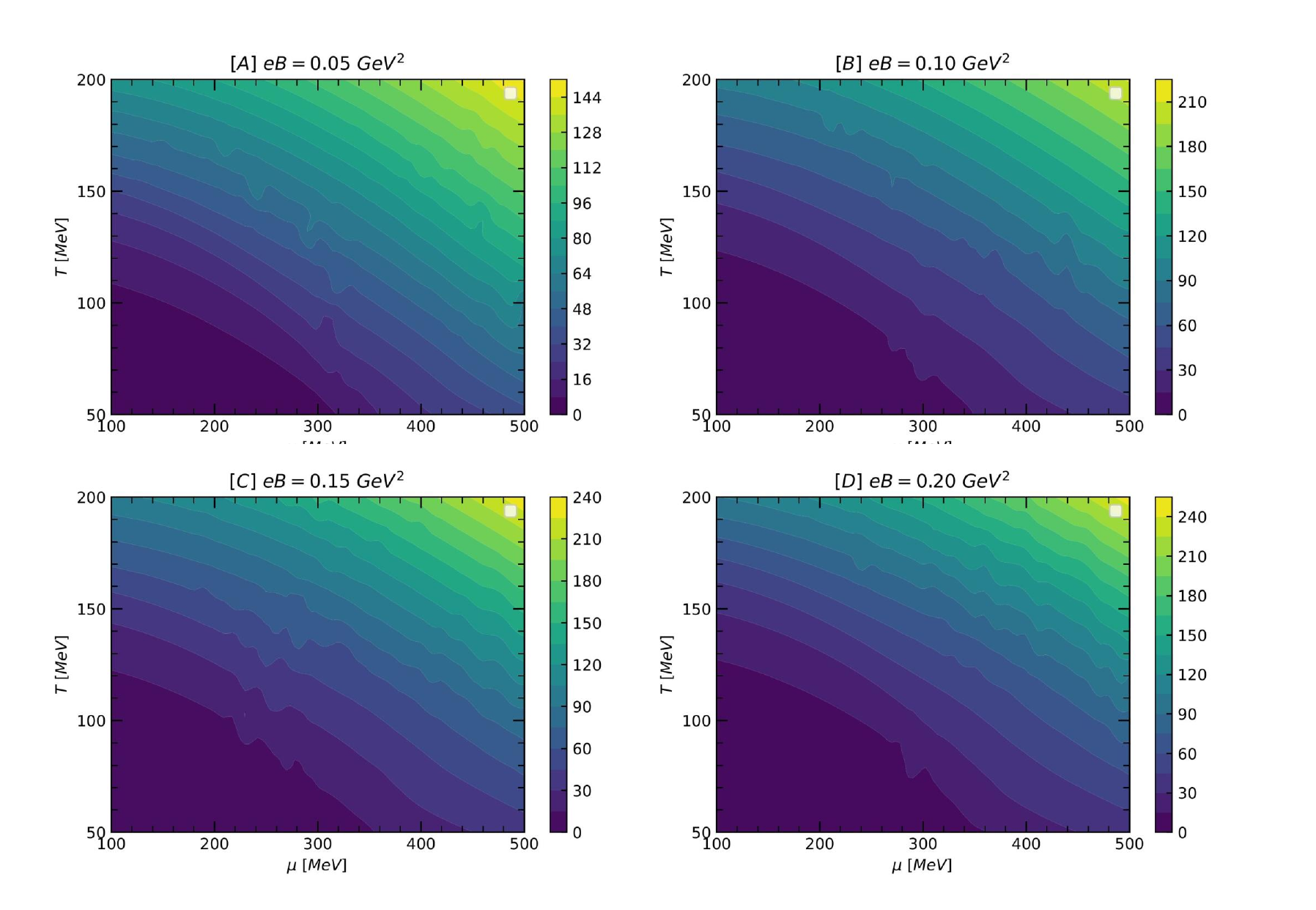}
	\caption{The contour maps of the entropy distributions in the T - $\mu$ plane. The panels correspond to different magnetic fields: A for $eB=0.05 \;\mathrm{GeV}^2$, B for $eB=0.10 \;\mathrm{GeV}^2$, C for $eB=0.15 \;\mathrm{GeV}^2$, and D for $eB=0.20 \;\mathrm{GeV}^2$.}
	\label{FIG.8}
\end{figure}

As a complementary study on the change of quark matter with the rise of temperature and chemical potential, the path for the transition from chiral condensation to tensor condensation also needs to be considered. The, isentropic diagrams have been plotted in Fig.\ref{FIG.8}. The color gradients depict the magnitude of entropy changes, with isentropic lines crossing both chiral and tensor condensation phases. The isentropic line as $S[M_f, F_f, B]~=~S[\mu, T, B]$ could connect tensor condensate with chiral restoration. In the coexistence of tensor and chiral condensation, $F_3$ oscillations always exist. Thermodynamic potential differences for different magnetic intensities between the point ($T = 60~\mathrm{MeV}$, $\mu = 470~\mathrm{MeV}$) and the point ($T = 100~\mathrm{MeV}$, $\mu = 160~\mathrm{MeV}$) is the transition curve of quark matter from low pressure and high temperature to high pressure and low temperature, and the thermodynamic potential density differences of $1~\mathrm{GeV}^3$ order. \par

	\section{SUMMARY AND CONCLUSIONS}\label{sec:summeraize}
	In this study, we have utilized the (2+1)-flavor NJL model to characterize the four-point interaction and investigated the relationship between chiral and tensor condensation under a strong magnetic field. We have identified the phase boundaries by varying the temperature and chemical potential. Our results indicate that the quark tensor condensate is primarily present in the low-temperature region and tends to diminish in the high-temperature region. Quark matter characterized by tensor condensation displays a magnetic catalytic effect. Additionally, the peak value of the tensor condensate is observed during the chiral restoration, and it gradually decreases as the temperature continues to increase. \par
	In our work, we have taken $F_8$ from the spin polarization as a measure of ferromagnetism and $F_3$ as a metric for spin density, which effectively illustrates the variations of the tensor condensate. Especially in the presence of a magnetic field $eB=0.05 \;\mathrm{GeV}^2$, the tensor condensation counteracts the energy loss associated with the dynamical loss of quark mass due to chiral symmetry restoration. Our analysis indicates that the spin polarization exhibits oscillations with the increasing of Landau energy levels. \par


\section{References}
	\begin{thebibliography}{99}
		\bibitem{Vachaspati:1991nm}
		T.~Vachaspati,
		"Magnetic fields from cosmological phase transitions,"
		Phys. Lett. B \textbf{265}, 258-261 (1991)

		\bibitem{Skokov:2009qp}
		V.~Skokov, A.~Y.~Illarionov and V.~Toneev,
		"Estimate of the magnetic field strength in heavy-ion collisions,"
		Int. J. Mod. Phys. A \textbf{24}, 5925-5932 (2009)

		\bibitem{Deng:2012pc}
		W.~T.~Deng and X.~G.~Huang,
		"Event-by-event generation of electromagnetic fields in heavy-ion collisions,"
		Phys. Rev. C \textbf{85}, 044907 (2012)
		
		\bibitem{Kurkela:2022elj}
		A.~Kurkela,
		"Thoughts about the utility of perturbative QCD in the cores of neutron stars \textendash{} contribution to a roundtable discussion on neutron stars and QCD,"
		EPJ Web Conf. \textbf{274}, 07008 (2022)
		
		\bibitem{CamaraPereira:2016chj}
		R.~C\^amara Pereira, P.~Costa and C.~Provid\^encia,
		"Two-solar-mass hybrid stars: a two model description with the Nambu-Jona-Lasinio quark model,"
		Phys. Rev. D \textbf{94}, no.9, 094001 (2016)
		
		\bibitem{Baym:2017whm}
		G.~Baym, T.~Hatsuda, T.~Kojo, P.~D.~Powell, Y.~Song and T.~Takatsuka,
		"From hadrons to quarks in neutron stars: a review,"
		Rept. Prog. Phys. \textbf{81}, no.5, 056902 (2018)
		
		\bibitem{Tsue:2012nx}
		Y.~Tsue, J.~da Providencia, C.~Providencia, M.~Yamamura and H.~Bohr,
		"Interplay between Spin Polarization and Color Superconductivity in High Density Quark Matter,"
		PTEP \textbf{2013}, no.10, 103D01 (2013)
		
		\bibitem{Maruyama:2017mqv}
		T.~Maruyama and T.~Tatsumi,
		"Spontaneous spin polarization due to tensor self-energies in quark matter,"
		Phys. Rev. D \textbf{96}, no.9, 096016 (2017)
		
		\bibitem{Maruyama:2000cw}
		T.~Maruyama and T.~Tatsumi,
		"Ferromagnetism of nuclear matter in the relativistic approach,"
		Nucl. Phys. A \textbf{693}, 710-730 (2001)
		
		\bibitem{Tsue:2016age}
		Y.~Tsue, J.~da Providencia, C.~Providencia, M.~Yamamura and H.~Bohr,
		"Spin polarization in high density quark matter under a strong external magnetic field,"
		Int. J. Mod. Phys. E \textbf{25}, no.12, 1650106 (2016)
		
		\bibitem{Tsue:2012jz}
		Y.~Tsue, J.~da Providencia, C.~Providencia and M.~Yamamura,
		"Effective Potential Approach to Quark Ferromagnetization in High Density Quark Matter,"
		Prog. Theor. Phys. \textbf{128}, 507-522 (2012)

		\bibitem{Fukushima:2010bq}
		K.~Fukushima and T.~Hatsuda,
		"The phase diagram of dense QCD,"
		Rept. Prog. Phys. \textbf{74}, 014001 (2011)

		\bibitem{Watts:2016uzu}
		A.~L.~Watts, N.~Andersson, D.~Chakrabarty, M.~Feroci, K.~Hebeler, G.~Israel, F.~K.~Lamb, M.~C.~Miller, S.~Morsink and F.~\"Ozel, \textit{et al.}
		"Colloquium : Measuring the neutron star equation of state using x-ray timing,"
		Rev. Mod. Phys. \textbf{88}, no.2, 021001 (2016)
		
		\bibitem{Riley:2019yda}
		T.~E.~Riley, A.~L.~Watts, S.~Bogdanov, P.~S.~Ray, R.~M.~Ludlam, S.~Guillot, Z.~Arzoumanian, C.~L.~Baker, A.~V.~Bilous and D.~Chakrabarty, \textit{et al.}
		"A $NICER$ View of PSR J0030+0451: Millisecond Pulsar Parameter Estimation,"
		Astrophys. J. Lett. \textbf{887}, no.1, L21 (2019)
		
		\bibitem{Ferrer:2010wz}
		E.~J.~Ferrer, V.~de la Incera, J.~P.~Keith, I.~Portillo and P.~L.~Springsteen,
		"Equation of State of a Dense and Magnetized Fermion System,"
		Phys. Rev. C \textbf{82}, 065802 (2010)

		\bibitem{Paulucci:2010uj}
		L.~Paulucci, E.~J.~Ferrer, V.~de la Incera and J.~E.~Horvath,
		"Equation of state for the MCFL phase and its implications for compact star models,"
		Phys. Rev. D \textbf{83}, 043009 (2011)

		\bibitem{Menezes:2009uc}
		D.~P.~Menezes, M.~Benghi Pinto, S.~S.~Avancini and C.~Providencia,
		"Quark matter under strong magnetic fields in the su(3) Nambu-Jona-Lasinio Model,"
		Phys. Rev. C \textbf{80}, 065805 (2009)

		\bibitem{Menezes:2008qt}
		D.~P.~Menezes, M.~Benghi Pinto, S.~S.~Avancini, A.~Perez Martinez and C.~Providencia,
		"Quark matter under strong magnetic fields in the Nambu-Jona-Lasinio Model,"
		Phys. Rev. C \textbf{79}, 035807 (2009)

		\bibitem{Avancini:2011zz}
		S.~S.~Avancini, D.~P.~Menezes and C.~Providencia,
		"Finite temperature quark matter under strong magnetic fields,"
		Phys. Rev. C \textbf{83}, 065805 (2011)

		\bibitem{Ferrer:2013noa}
		E.~J.~Ferrer, V.~de la Incera, I.~Portillo and M.~Quiroz,
		"New look at the QCD ground state in a magnetic field,"
		Phys. Rev. D \textbf{89}, no.8, 085034 (2014)

		\bibitem{tHooft:1976snw}
		G.~'t Hooft,
		"Computation of the Quantum Effects Due to a Four-Dimensional Pseudoparticle,"
		Phys. Rev. D \textbf{14}, 3432-3450 (1976)
		[erratum: Phys. Rev. D \textbf{18}, 2199 (1978)]

		\bibitem{Koch:1987py}
		V.~Koch, T.~S.~Biro, J.~Kunz and U.~Mosel,
		"A Chirally Invariant Fermionic Field Theory for Nuclear Matter,"
		Phys. Lett. B \textbf{185}, 1-5 (1987)

		\bibitem{Niembro:1990tc}
		R.~Niembro, S.~Marcos, M.~L.~Quelle and J.~Navarro,
		"Magnetic susceptibility of neutron matter in a relativistic approach,"
		Phys. Lett. B \textbf{249}, 373-376 (1990)

		\bibitem{Bohr:2012mg}
		H.~Bohr, P.~K.~Panda, C.~Providencia and J.~da Providencia,
		"Ferromagnetic condensation in high density hadronic matter,"
		[arXiv:1203.6272 [nucl-th]].

		\bibitem{Tatsumi:1999ab}
		T.~Tatsumi,
		"Ferromagnetism of quark liquid,"
		Phys. Lett. B \textbf{489}, 280-286 (2000)

		\bibitem{Kagawa:2020vsf}
		A.~Kagawa, M.~Morimoto, Y.~Tsue, J.~da Provid\^encia, C.~Provid\^encia and M.~Yamamura,
		"Nonzero tensor condensates in cold quark matter within the three-flavor Nambu\textendash{}Jona\textendash{}Lasinio model with the Kobayashi\textendash{}Maskawa\textendash{}\textquoteright{}t Hooft interaction,"
		Int. J. Mod. Phys. E \textbf{29}, no.06, 2050036 (2020)

		\bibitem{Qiu:2023kwv}
		Y.~W.~Qiu and S.~Q.~Feng,
		"Spin polarization and anomalous magnetic moment in a (2+1)-flavor Nambu\textendash{}Jona-Lasinio model in a thermomagnetic background,"
		Phys. Rev. D \textbf{107}, no.7, 076004 (2023)

		\bibitem{Bao:2024glw}
		Y.~R.~Bao and S.~Q.~Feng,
		"Effects of tensor spin polarization on the chiral restoration and deconfinement phase transitions,"
		Phys. Rev. D \textbf{109}, no.9, 096033 (2024)

		\bibitem{Lin:2022ied}
		F.~Lin, K.~Xu and M.~Huang,
		"Magnetism of QCD matter and the pion mass from tensor-type spin polarization and the anomalous magnetic moment of quarks,"
		Phys. Rev. D \textbf{106}, no.1, 016005 (2022)

		\bibitem{Buballa:1996tm}
		M.~Buballa,
		"The Problem of matter stability in the Nambu-Jona-Lasinio model,"
		Nucl. Phys. A \textbf{611}, 393-408 (1996)

		\bibitem{Abhishek:2018usv}
		A.~Abhishek, A.~Das, H.~Mishra and R.~K.~Mohapatra,
		"Spin Polarization and Chiral Condensation in 2+1 flavor Nambu-Jona-Lasinio model at finite temperature and baryon chemical potential,"
		Phys. Rev. D \textbf{100}, no.11, 114012 (2019)

		\bibitem{Buballa:2003qv}
		M.~Buballa,
		"NJL model analysis of quark matter at large density,"
		Phys. Rept. \textbf{407}, 205-376 (2005)

		\bibitem{Noronha:2007wg}
		J.~L.~Noronha and I.~A.~Shovkovy,
		"Color-flavor locked superconductor in a magnetic field,"
		Phys. Rev. D \textbf{76}, 105030 (2007)

		\bibitem{Fraga:2008qn}
		E.~S.~Fraga and A.~J.~Mizher,
		"Chiral transition in a strong magnetic background,"
		Phys. Rev. D \textbf{78}, 025016 (2008)

		\bibitem{Klevansky:1992qe}
		S.~P.~Klevansky,
		"The Nambu-Jona-Lasinio model of quantum chromodynamics,"
		Rev. Mod. Phys. \textbf{64}, 649-708 (1992)

		\bibitem{Kohyama:2016fif}
		H.~Kohyama, D.~Kimura and T.~Inagaki,
		"Parameter fitting in three-flavor Nambu\textendash{}Jona-Lasinio model with various regularizations,"
		Nucl. Phys. B \textbf{906}, 524-548 (2016)

		\bibitem{Matsuoka:2018huq}
		H.~Matsuoka, Y.~Tsue, J.~Da Provid\^encia, C.~Provid\^encia and M.~Yamamura,
		"Hybrid stars from the NJL model with a tensor-interaction,"
		Phys. Rev. D \textbf{98}, no.7, 074027 (2018)

		\bibitem{Kohyama:2015hix}
		H.~Kohyama, D.~Kimura and T.~Inagaki,
		"Regularization dependence on phase diagram in Nambu\textendash{}Jona-Lasinio model,"
		Nucl. Phys. B \textbf{896}, 682-715 (2015)

		\bibitem{Hatsuda:1994pi}
		T.~Hatsuda and T.~Kunihiro,
		"QCD phenomenology based on a chiral effective Lagrangian,"
		Phys. Rept. \textbf{247}, 221-367 (1994)

		\bibitem{Feng:2011fj}
		B.~Feng, E.~J.~Ferrer and V.~de la Incera,
		"Cooper Pair's Magnetic Moment in MCFL Color Superconductivity,"
		Nucl. Phys. B \textbf{853}, 213-239 (2011)

		\bibitem{Pagura:2016pwr}
		V.~P.~Pagura, D.~Gomez Dumm, S.~Noguera and N.~N.~Scoccola,
		"Magnetic catalysis and inverse magnetic catalysis in nonlocal chiral quark models,"
		Phys. Rev. D \textbf{95}, no.3, 034013 (2017)

		\bibitem{Bali:2012zg}
		G.~S.~Bali, F.~Bruckmann, G.~Endrodi, Z.~Fodor, S.~D.~Katz and A.~Schafer,
		"QCD quark condensate in external magnetic fields,"
		Phys. Rev. D \textbf{86}, 071502 (2012)

		\bibitem{DElia:2018xwo}
		M.~D'Elia, F.~Manigrasso, F.~Negro and F.~Sanfilippo,
		"QCD phase diagram in a magnetic background for different values of the pion mass,"
		Phys. Rev. D \textbf{98}, no.5, 054509 (2018)
		
		
		
	\end{thebibliography}
\end{document}